\documentclass[aps,prl]{revtex4}
\def\ee{\end{equation}}
\def\bea{\begin{eqnarray}}

\usepackage{graphicx}
\DeclareGraphicsExtensions{.png}
\begin{document}

\title{Stronger Tests of the Collapse Locality Loophole in Bell Experiments}
\author{ Adrian Kent}
\email{A.P.A.Kent@damtp.cam.ac.uk} 
\affiliation{Centre for Quantum Information and Foundations, DAMTP, Centre for
Mathematical Sciences,
University of Cambridge, Wilberforce Road, Cambridge, CB3 0WA, United Kingdom}
\affiliation{Perimeter Institute for Theoretical Physics, 31 Caroline Street
North, Waterloo, ON N2L 2Y5, Canada.}

\begin{abstract}
Several versions of quantum theory assume some form of localized
collapse.   If measurement outcomes are indeed defined by
localized collapses, then a loophole-free demonstration of 
Bell non-locality needs to ensure space-like separated collapses 
associated with the measurements of the entangled systems.
This collapse locality loophole remains largely untested, 
with one significant exception probing Diosi's and Penrose's gravitationally
induced collapse hypotheses.
I distinguish two versions of the
loophole, and focus on the more strongly motivated version,
in which local hidden variables may depend causally on collapse
outcomes but may not independently depend on measurement settings. 
I describe here techniques that allow much stronger
experimental tests. 
These apply to all the well known types of collapse postulate,
including gravitationally induced collapse, spontaneous 
localization models and Wigner's consciousness-induced collapse.    
\end{abstract}
\maketitle
\section{Introduction}

Bell's work \cite{bell1966js,bell1989einstein} on the empirical 
implications of the Einstein--Podolsky--Rosen argument \cite{einstein1935can}
showed that quantum theory violates local causality \cite{bell1976theory}.  
It has led to many beautiful experiments 
that aim to refute local hidden variable theories (LHVT). 
To eliminate as many extra assumptions as possible, and
so refute as wide a class of LHVT as possible, 
experiments have been devised that close or minimize the
detector efficiency 
\cite{Pearle70,RKMSIMW01,MMMOM08,GMRWKBLCGNUZ13} 
and locality and
setting-independence loopholes \cite{ADR82,WJSWZ98,TBZG98,GZ99}.  
Analysis of the memory loophole showed that it too can be closed 
by using appropriate statistical tests in standard Bell experiments
\cite{barrett2002quantum,gill2003accardi}.

More recently, what were described ``loophole-free'' experiments, closing
all three loopholes simultaneously, were announced
\cite{hensen2015loophole,shalm2015strong,giustina2015significant}. 
While these were undoubtedly impressive experiments that achieved 
a long-sought goal, their definitiveness was overstated.
Significant and theoretically interesting loopholes remain. 
In particular, the collapse locality loophole \cite{kent2005causal} remains
largely untested.   This loophole arises because, while Bell experiments are
supposed to demonstrate non-local correlations between measurement
outcomes on spacelike separated systems, we do not know for sure
where in spacetime the relevant measurement outcomes actually arise.
In some versions of quantum theory this question does not have a well-defined
answer.  However, in versions in which collapse is an objective and
localized process, it does.   There is not a consensus among theorists
that objective collapse hypotheses with localized collapses are
necessary, and they may not be correct.   However, there are a variety of good motivations for taking 
them seriously, given the problems in making sense of unitary quantum
theory \cite{saunders2010many} and quantum gravity \cite{sep-quantum-gravity}.  
It is questionable whether any Bell experiment to date has created
spacelike separated measurement outcomes, 
according to most well-known collapse 
hypotheses
(e.g. \cite{wigner1995remarks,ghirardi1986unified,ghirardi1990markov}). 

For example, the Ghirardi-Rimini-Weber dynamical collapse model\cite{ghirardi1986unified} requires  
a measurement-type interaction to 
create a superposition of distinguishable position states for a 
large number of correlated particles in order to produce a
significant probability of a definite measurement event.
Similarly, mass-dependent continuous spontaneous localization
models\cite{pearle1989combining,ghirardi1990markov,
pearle1994bound}
require the creation of a superposition of significantly
different mass distributions in order to produce a
significant probability of a definite measurement event.
The photodetector avalanches generated by photon measurements
in many Bell experiments do not involve sufficiently many
particles, and the states of wires carrying currents
transmitting the results are not sufficiently distinguishable, 
according to the criteria these models define.   
Of course, dynamical collapse models do predict that collapses 
defining measurement outcomes eventually take place, but we expect this to 
be later in the causal chain,
when the results are correlated and stored in computer
memory, printed out, or read by experimenters. 
There is no guarantee that such collapses are spacelike separated.
Indeed, only one Bell experiment to date \cite{SBHGZ08} has been
specifically designed to ensure that, given a specific 
objective collapse hypothesis (due to Penrose and Diosi),
spacelike separated collapses define the measurement outcomes
in the two wings.   We consider this experiment, and its
relation to the collapse hypotheses of Refs. \cite{ghirardi1986unified,pearle1989combining,ghirardi1990markov,
pearle1994bound}, in more
detail below. 

Tests of the collapse locality loophole can be motivated as 
tests of standard quantum theory 
against a class of apparently internally consistent, albeit
strange, alternatives, collectively termed causal quantum
theory \cite{kent2005causal,kent2018testing}.  
However, the loophole {\it per se} does not logically rely
on the consistency or plausibility of 
causal quantum theory.    Testing it tests quantum 
theory against the general class of local hidden variable theories
in which collapses, and thus measurement outcomes, causally influence
and may be causally influenced by the local hidden variables.
We will assume this more general motivation here. 
We are interested in testing the hypothesis that spacelike
separated measurement outcomes respect Bell inequalities. 
This is consistent with the outcomes of pairs of measurements
that are simultaneous and co-located, or lightlike or timelike
separated, respecting standard quantum
predictions (and thus violating Bell inequalities).
We need to assume this in order to explain the results
of Bell experiments to date.  

Ideally, one would ultimately like to close
as many loopholes as possible in a single experiment:
hence the motivation for experiments such as
those in Refs. \cite{hensen2015loophole,shalm2015strong,giustina2015significant}. 
However, this involves tradeoffs, and it may not yet 
be possible to close all plausible loopholes simultaneously. 
For this reason, many pre-2015 Bell experiments
were designed to close only a single loophole, or subset
of loopholes.   
One motivation for addressing loopholes separately
is the intuition that we should
assign a relatively small Bayesian prior to the hypothesis
that nature exploits any given unclosed loophole, and
a much smaller prior probability to the hypothesis that nature 
exploits two or more loopholes simultaneously in a way 
that is only evident when both or all apply.   Put another way,
while hidden variable theories that exploit one loophole might 
just possibly have some theoretical motivation, 
those that require two or more in combination seem
far more conspiratorial and implausible.    

This justification seems particularly strong in the case of 
the collapse locality loophole.
As noted above, there are theoretical motivations for considering
the hypothesis that localized collapses play a fundamental 
role in physics, and if what we call measurements are fundamentally
defined by localized collapses then we really need to 
close the collapse locality loophole in order to demonstrate
Bell non-local correlations between measurement outcomes.
It is not immediately obvious that this line of thought adds any particular
motivation to hidden variable theories that might exploit
other loopholes.  
This motivates us to 
consider experiments that test the collapse
locality loophole as strongly as possible given
current technology, without considering other
loopholes, and we focus on such experiments in this paper. 

\begin{figure}[h]
\centering
\includegraphics[width=\linewidth, height=10cm]{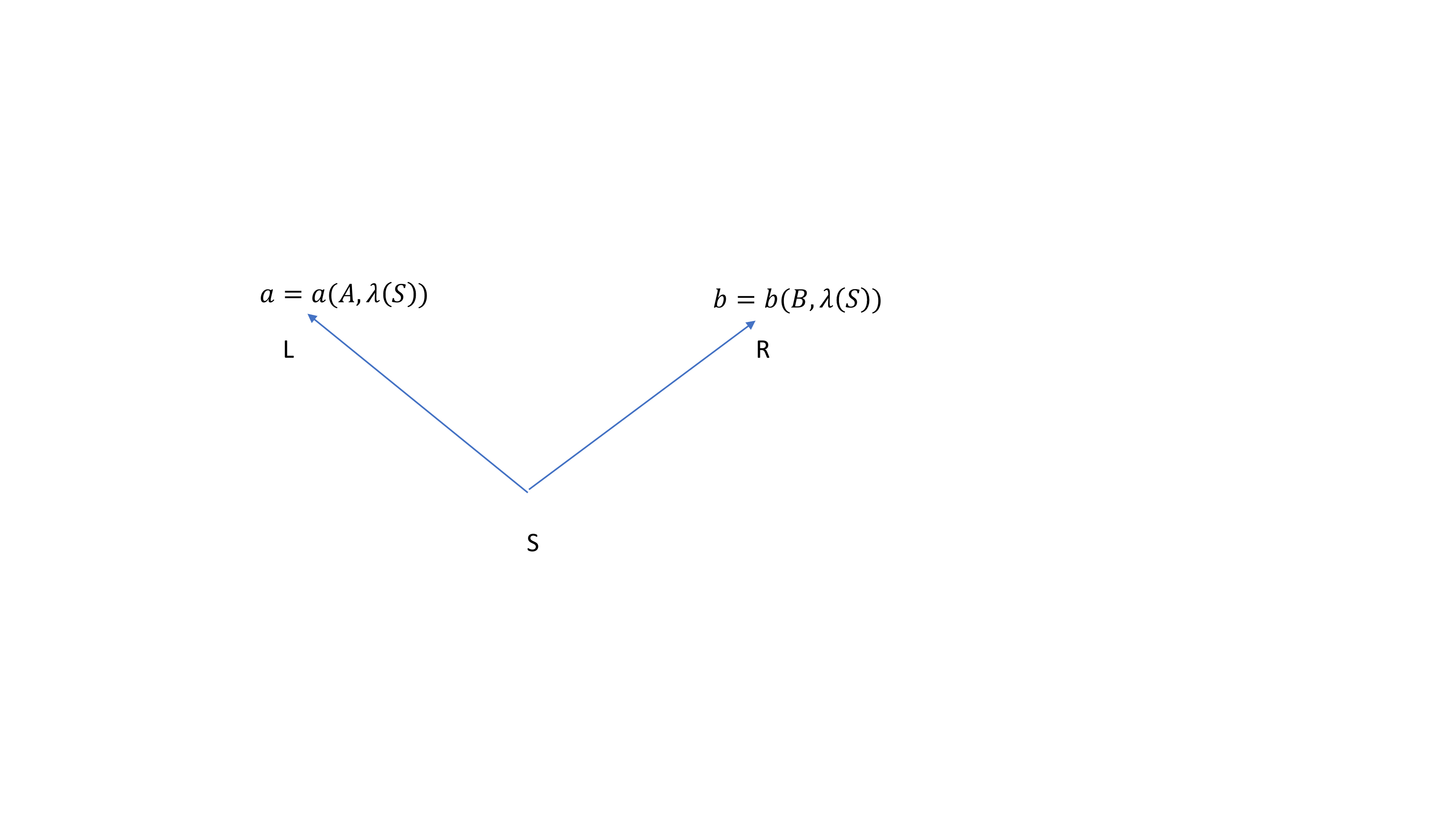} 
\caption{Schematic description of a standard Bell experiment}
\label{one}
\end{figure}

In the standard framework for describing Bell experiments, the
collapse locality loophole can be understood as follows. 
In a schematic description of a standard Bell experiment
(Fig. \ref{one}) a source S generates entangled particles
which propagate to devices in wings L and R.
Measurement setting choices $A$ and $B$ in the two wings
are made locally, producing outcomes $a$ and $b$ respectively.
A local hidden variable theory would allow these outcomes
to depend on a common local hidden variable $\lambda$, which
depends on events at the source S and in its past light cone.   In principle, the
local hidden variables at L may depend on other events
in its past light cone (and similarly R).   
We make here the common assumption that the relevant events in the past 
light cone of L (including those in the past light cone of S) 
are effectively uncorrelated with the measurement choice
B, and similarly R and A, excluding superdeterminist explanations
for Bell correlations.  To simplify the notation we list explicitly
only the dependences crucial for our discussion; thus we write
$\lambda = \lambda(S)$ to emphasize its dependence on events at
and in the past of S.

\begin{figure}[h]
\centering
\includegraphics[width=\linewidth, height=10cm]{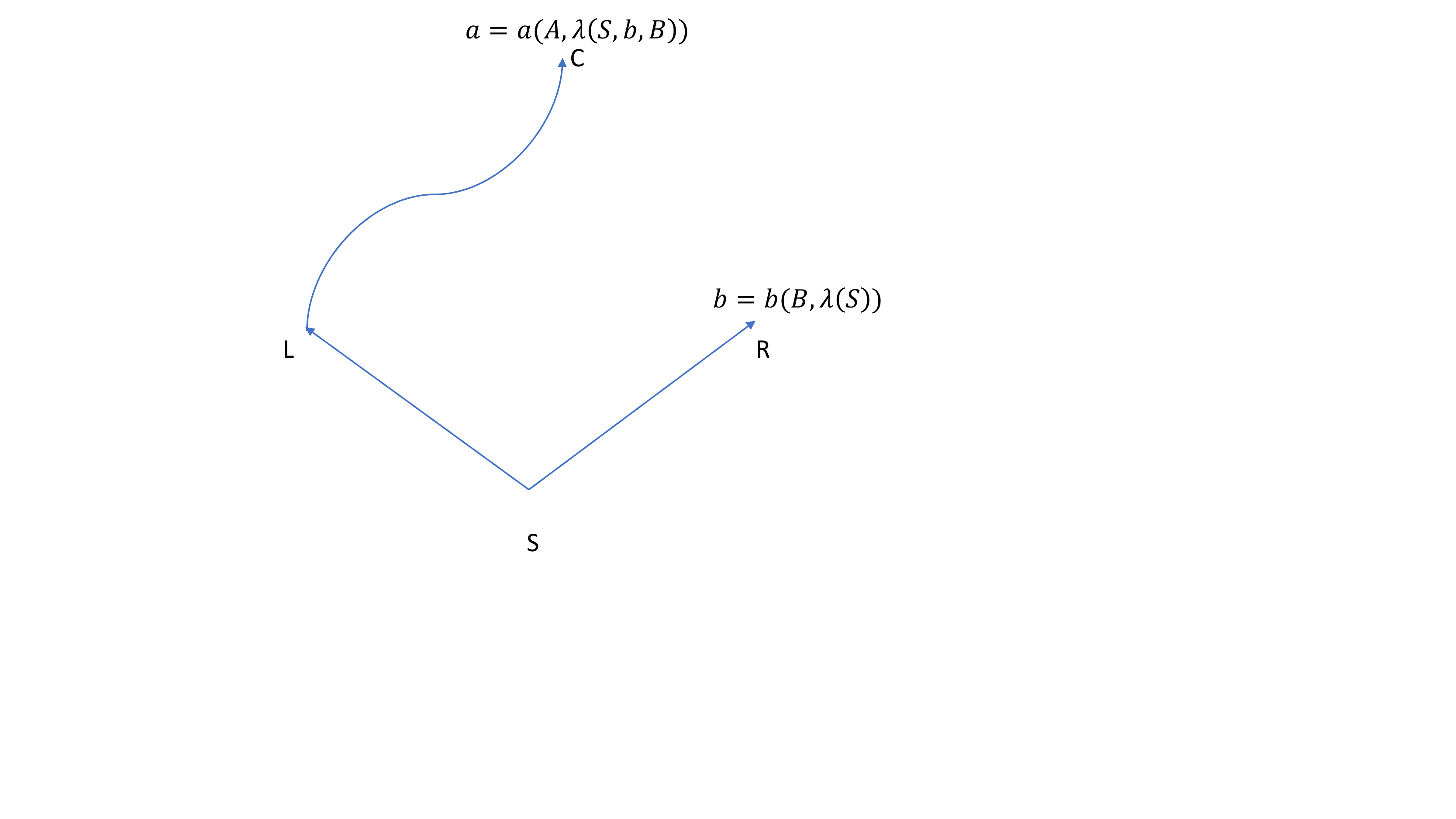} 
\caption{A Bell experiment in which the collapse locality loophole
allows records of outcomes, apparently from one wing of the experiment
but actually generated at a point in the causal future of both, to depend
on measurement settings and outcomes in the other wing.
If the outcomes are indeed generated by collapse events at R, then both versions of 
the collapse locality loophole allow the outcomes at C to depend on the settings
and outcomes at R.}
\label{two}
\end{figure}

Now consider the possibility that the outcome of measurement 
A is actually determined by a collapse event at a point C in the causal
future of the collapse event determining the outcome of 
measurement $B$ (Fig. \ref{two}). 
In this case the local hidden variables at C may depend explicitly    
on this prior collapse event, and hence on
the choice $B$ and outcome $b$ as well as on events at S. 

We need to distinguish here between two hypotheses associated
with collapse and locality in Bell experiments.
Since these have not previously been clearly separated, 
we propose to refine the terminology.   

In principle, the local hidden variables at C could depend
on everything in their past light cone.  In particular,
they could depend on the measurement choices $B$, simply
because a configuration has been chosen for 
the measurement apparatus at R and this physical
fact {\it per se} may influence variables in the future light cone
in a general local hidden variable theory.   
An explanation of this type effectively relies on the locality
loophole, together with the hypothesis that the relevant 
measurement event actually takes place at a point (C) timelike separated
from the point (R) where the measurement settings were made, 
even though the experiment was intended
to make measurements at spacelike separated points (L and R). 
We refer to this as exploiting the {\it extended collapse
locality loophole}.  

A more restrictive hypothesis is that the local hidden
variables depend significantly on the choice $B$ and outcome $b$  
only if a collapse event indeed took place at R.
On this view, simply setting up the measurement apparatus at R
so as to define a particular choice of measurement does not {\it per se} 
have a relevant effect -- an effect that significantly alters the predicted
correlations -- on the local hidden variables at C, even though C
is in the future light cone of R.   
If the measurement at R does not produce a
collapse event there, then the local hidden variables at C 
may be treated as though effectively independent of the measurement
settings.  
However, a collapse of the subsystem at R is a physical
event that may affect the local hidden variables of the 
entangled subsystem at C, in such a way as to alter
the predicted correlations.   According to 
quantum theory, a collapse at R with outcome $b$ 
projects the state of the local subsystem onto 
the eigenstate with eigenvalue $b$ of the observable
defined by measurement choice $B$, a state which depends
on both $B$ and $b$.   The hypothesis is that such
collapse events significantly affect the local hidden
variable dynamics, so that the local hidden variables
in the causal future of R may depend on both $B$ and $b$
in a way that affects the observed correlations. 
In particular, since C is in the future light cone of R, 
the local hidden variables there may depend on both $B$ and $b$. 
We refer to this as exploiting the {\it essential collapse locality loophole}. 

Our motivation for this second hypothesis, and for
the term ``essential'', comes from applying 
EPR's \cite{einstein1935can} and Bell's \cite{bell1966js,bell1989einstein,bell1976theory}
arguments to models where collapses are physically objective and localized.
EPR's discussion of measurements on entangled
systems suggested that quantum theory may be 
incomplete, since -- given premises that are arguable, although of
course presently generally rejected -- a measurement
on one subsystem seems to change the physical properties of 
a distant subsystem.  Bell's discussion of locally causal hidden
variables suggests a natural way in which more complete
underlying theories, consistent with Einstein causality, could be defined. 
Theories with objective localized collapses offer a precise
definition of measurement, and hence of 
the events to which the EPR argument should apply.
They also introduce localized physical events (the collapses)
and locally created physical data (the collapsed states)
that supplement and modify unitary quantum dynamics.   
A local hidden variable theory underlying
an objective localized collapse version of
quantum theory thus ought to allow the distribution
of local hidden variables for a system 
to depend causally on collapse data associated with
that system as well as on its initial
quantum state.  
  
\begin{figure}[h]
\centering
\includegraphics[width=\linewidth, height=10cm]{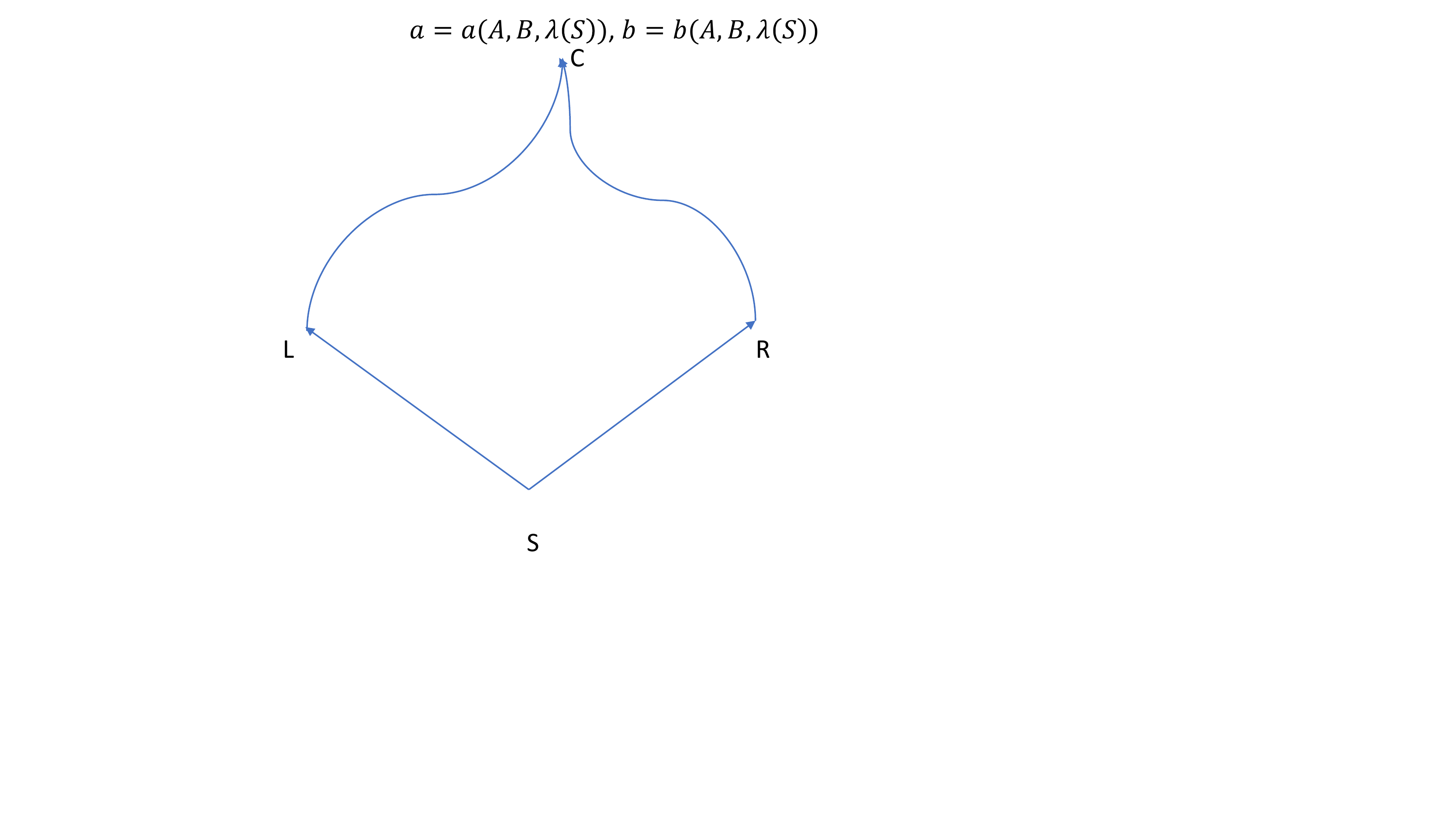} 
\caption{Another application of the collapse locality loophole.
Here the records of outcomes, apparently from both wings, are
generated together at point C.  The recorded measurement outcomes
for each wing may depend on both measurement settings.}
\label{three}
\end{figure}

Another possibility (Fig. \ref{three}) is that the outcomes of
measurements $A$ and $B$, apparently obtained at L and R, are actually jointly determined
by co-located collapse events at C. 
In this case the local hidden variables
at C determining both outcomes ($a$ and $b$)
may depend explicitly on both measurement settings (A and B).
This hypothesis may be justified if the detectors at L and R are similar, and the detector readings are
transmitted along similar channels to a device at C that stores them
and calculates correlations. 
With this set up, an objective localized collapse theory will generally
either predict collapses in the vicinity of both L and R (either in
the detectors or at an early stage within the communication channels) or neither.
If neither, then it may predict no 
collapse until the quantum state recording the outcome data
undergoes appropriate further evolution and/or interactions.
The qualitative and quantitative features of the evolution and interactions required
to induce collapses are determined by the specifics of the collapse
theory.
The essential and extended versions of the collapse locality loophole
are hard to separate in this case, since the localized collapse theory may imply that
there is effectively a single joint measurement of $A$ and $B$ at C. 

In either of the configurations described in Figs. \ref{two} and \ref{three}, local hidden variable theories
can reproduce the predictions of quantum theory, violating Bell
inequalities and producing ``faux non-local'' correlations, 
which appear to verify Bell non-locality but actually do not.  

Discussion of the collapse locality loophole \cite{kent2005causal}
motivated a beautiful 
experiment by Salart et al.\cite{SBHGZ08}, 
which tested the loophole for the specific collapse 
hypotheses sketched by Diosi \cite{diosi1987universal} and Penrose \cite{penrose1996gravity}, according 
to which wave function collapse takes place to 
prevent a superposition of distinguishable gravitational fields. 
Diosi and Penrose proposed quantitative estimates for the 
distinguishability of mass distributions in superposition
components required for collapse, and Salart et al. were
able to arrange a configuration of piezocrystals coupled
to a Bell experiment so that, on Diosi and Penrose's
estimates, the relevant collapses would indeed be
spacelike separated.

The Salart et al. experiments confirmed the predictions
of quantum theory, thus closing the loophole under the
precise assumptions made.  
That said, Diosi and Penrose's estimates were based on 
heuristic calculations rather than derived from a 
consistent underlying theory, and altering them by
a factor of $\approx 10^2$ would leave the loophole
open in the Salart et al. experiment.  
The experimental analysis
did not address other hypotheses about gravitationally induced collapse,
or other types of spontaneous collapse models (e.g. \cite{ghirardi1986unified,ghirardi1990markov}).
Nor, of course, did the experiment address Wigner's speculative 
(but motivated) hypothesis \cite{wigner1995remarks} that measurement outcomes
and collapse might require conscious observation. 

There thus remains very strong motivation for stronger, 
more systematic and more general tests of both forms of the collapse
locality loophole.  In this paper I describe simple techniques
that enable some tests of the essential collapse locality loophole. 

\subsection{Empirical tests: Bell experiments}

Non-local correlations appear to have been demonstrated in 
many Bell experiments.  However, because of the collapse
locality loophole, appearances may possibly have been deceptive.
Consider again a typical Bell experiment involving an entangled
pair of photons, generated by a source S, whose polarizations are measured by 
a combination of filters and photodetectors in local labs
at space-like separated points L and R, following the layout of
Fig. \ref{three}.   

Speaking loosely -- in the way that physicists do when 
discussing quantum experiments when conceptual details
are not relevant -- a photon entering, say, lab L enters
a photodetector, generates an avalanche of photons, 
which generates a small electrical signal.   
Similar events take place in lab R.  The signals
are brought together and compared, producing 
a record of the results, and allowing their
correlations to be obtained, at some point C in
the joint future of L and R.   

The point at issue, in the context of the collapse
locality loophole, is where and when physical measurement(s) and collapse(s) 
actually take(s) place in such experiments.   
Is amplification of a single photon state to a
larger number of photons is sufficient to define a measurement?
Or generating a small electrical pulse from the photodetector?
According to most well-known collapse hypotheses, the answers
are no, or at least not necessarily. 
The relevant collapses and measurement outcomes may
instead have been co-located, at or after C.
For example, they may take place when the measurement
outcomes were amplified and recorded on a hard drive,
or when the printout was read by an experimenter. 

In the experiment mentioned above, Salart et al. \cite{SBHGZ08}
sent the electrical signals generated by the photomultipliers
directly through piezocrystals on each wing,
causing the piezocrystals to deform.   These deformations
were observed locally by interferometry, using mirrors
attached to one end of each piezocrystal.   The 
deformation of the piezocrystals and consequent movement
of the mirrors involves a relatively
large mass displacement, far larger than that created
by the photomultiplier avalanche or the electrical pulses.
The experimental parameters were chosen so that, according to Diosi and Penrose's
intutions and estimates, the two possible states of the piezocrystal (deformed
and undeformed) create macroscopically distinct
gravitational fields, which quickly causes a collapse when
the experiment would (according to unitary quantum
dynamics) place them in superposition.    

Salart et al.'s experiment verifed that
the piezoactuator displaced a $2$mg mirror of dimensions
$3 \times 2 \times 0.15$mm through a distance of 
$\geq 12.6$nm within $\approx 6$ $\mu{\rm s}$ of the photon
entering the analyzer.   
In collapse models, until collapse takes place,
measurement processes such as those in the experiment produce
superposition states.  The relevant superposition here is
of the undisplaced and displaced states.   
Salart et al. obtain a lower bound on the collapse time
of such a superposition by neglecting the actuator displacement
and considering the state of the mirror, which is effectively
in a superposition of two overlapping position states.   
This gives \cite{adler2007comments} a collapse time of $\leq \approx1 \mu{\rm s}$,
using Diosi's estimate (Penrose's estimate is a factor of two
smaller), and hence an upper bound of 
$\approx 7 \mu{\rm s}$ between the photon entering the
analyzer and a collapse event.   

The two
wings in the Salart et al. experiment were separated
by $\approx 60\mu{\rm s} \times c$, so that under the
stated assumptions the introduction of the piezocrystals
ensures spacelike separated collapses in the two wings. 
Without the piezocrystals, according to Diosi-Penrose, 
there would be no space-like separated collapses, since
the mass displacements due to the photomultiplier avalanches
and electrical pulses are negligible over the experimental
timescales.   

Salart et al. thus successfully closed the essential collapse locality 
loophole for gravitationally induced collapse, assuming 
that Diosi-Penrose's collapse time estimates are accurate.
It should be noted, though, that these estimates are based
on intuition and require some rather ad hoc assumptions, and
that they are not known to follow from a consistent dynamical theory
of gravitationally induced collapse.    
Somewhat lower collapse time bounds could be 
derived by allowing for the difference in densities
between the piezoactuator and mirror and modelling
the mass distribution of both.  
Still, it is unclear that the experiment would close the loophole
given an extra factor of $10^2$ or so in the Diosi-Penrose estimates. 
This already gives motivation for seeking stronger experiments. 

\subsection{Other collapse models}

It is interesting to consider the collapse times predicted in 
the Salart et al. experiments by other well known collapse
models.   The most extensively studied, and arguably the
best motivated, of these is the mass-dependent version 
of the continuous spontaneous localization model due to Ghirardi, Pearle
and Rimini \cite{pearle1989combining,ghirardi1990markov,
  pearle1994bound}.   
We make the same simplifying assumption as Salart et al., by 
considering only the state of the mirror, which before
collapse is in a superposition of overlapping position
states.   An analysis of the collapse rate \footnote{Relevant
approximations and calculations are set out in detail in Pearle's 
contribution to a forthcoming book \cite{pearle2019dynamical}.
I thank Philip Pearle for an advance copy.}
gives a collapse time of 
\begin{equation}
\frac{1}{4 \pi \lambda a^2} N^{-2} A \, ,
\end{equation}
where $\lambda$ and $a$ are the free parameters in the CSL
model, $N$ is the number of nucleons in the
sliver of the mirror that does not overlap both
superposition states, and $A$ is the area of the  
mirror surface.
For the parameter choices often used in example discussions,
$\lambda = 10^{-16} {\rm s}^{-1}$, $a = 10^{-5} {\rm cm}$, this
gives a collapse time of $\approx 10^{-8}$s, faster than the Diosi-Penrose
estimate.    

However, it is suggested in the literature \cite{aicardi1991dynamical,pearle2019dynamical,bassi2010breaking} that a value of $\lambda$ as
low as $ 10^{-19} {\rm s}^{-1}$ suffices to produce collapses fast
enough to be consistent with human perceptions of definite events,
which are the only certain data from which lower bounds can be
derived.  There are significant uncertainties in the derivation of 
this estimate, which may be too low to ensure that photon observations
generally cause collapses within the human eye (see e.g. \cite{kent2018perception}).
On the other hand, one can argue \cite{pearle2019dynamical,kent2018perception} that observers generally
produce physiological responses to any significant observation, 
and that a collapse model may need only to ensure that collapses
take place very rapidly after such a response.    
If this is accepted, values of $\lambda$ significantly lower than 
$10^{-19}$ may be consistent with our perceptions.
Indeed, if the relevant physiological responses are essentially always
macroscopic -- say, involving at least $1$g of body matter moving at
least $1$mm -- then a firm lower bound would be many orders
of magnitude lower.    It is thus hard to argue that 
the Salart et al. experiment has definitely closed the CSL
version of the essential collapse locality loophole.    

Similar comments apply to the earlier Ghirardi-Rimini-Weber (GRW)
collapse model.   This model, in its original version, requires
us to treat all the nucleons in the relevant system as distinguishable
particles, each of which has the same collapse rate.
It is thus natural to consider the piezocrystal
and mirror as one system.
We can simplify the estimate by taking the two superposition
states to be displacements of the entire system by 
the average value $\approx \frac{d}{2}$ (in the actual
states, the nucleon displacements range from $0$ to $d$).
This gives a GRW collapse time estimate of 
\begin{equation}
\frac{16 a^2}{\lambda N d^2 } \, 
\end{equation}
where $\lambda$ and $a$ are GRW model paramters, $d$ is the
displacement and $N$ the number of nucleons in the piezocrystal
and mirror. 
For the often used example values $\lambda = 10^{-16}{\rm s}^{-1}$ and
$a = 10^{-5}$cm, this gives a collapse time of $\approx 2 \times
10^{-4}$s.   The Salart et al. experiment thus did not close
the GRW version of the essential collapse locality loophole
even with these standard example parameters.   Finding a firmly defensible lower
bound for the parameter $\lambda$ is as problematic for
GRW models as for CSL models, so that the Salart et al. 
experiment is far from closing the GRW version of the 
essential collapse locality loophole.   

\subsection{Extended experiments} 

\begin{figure}[h]
\centering
\includegraphics[width=\linewidth, height=10cm]{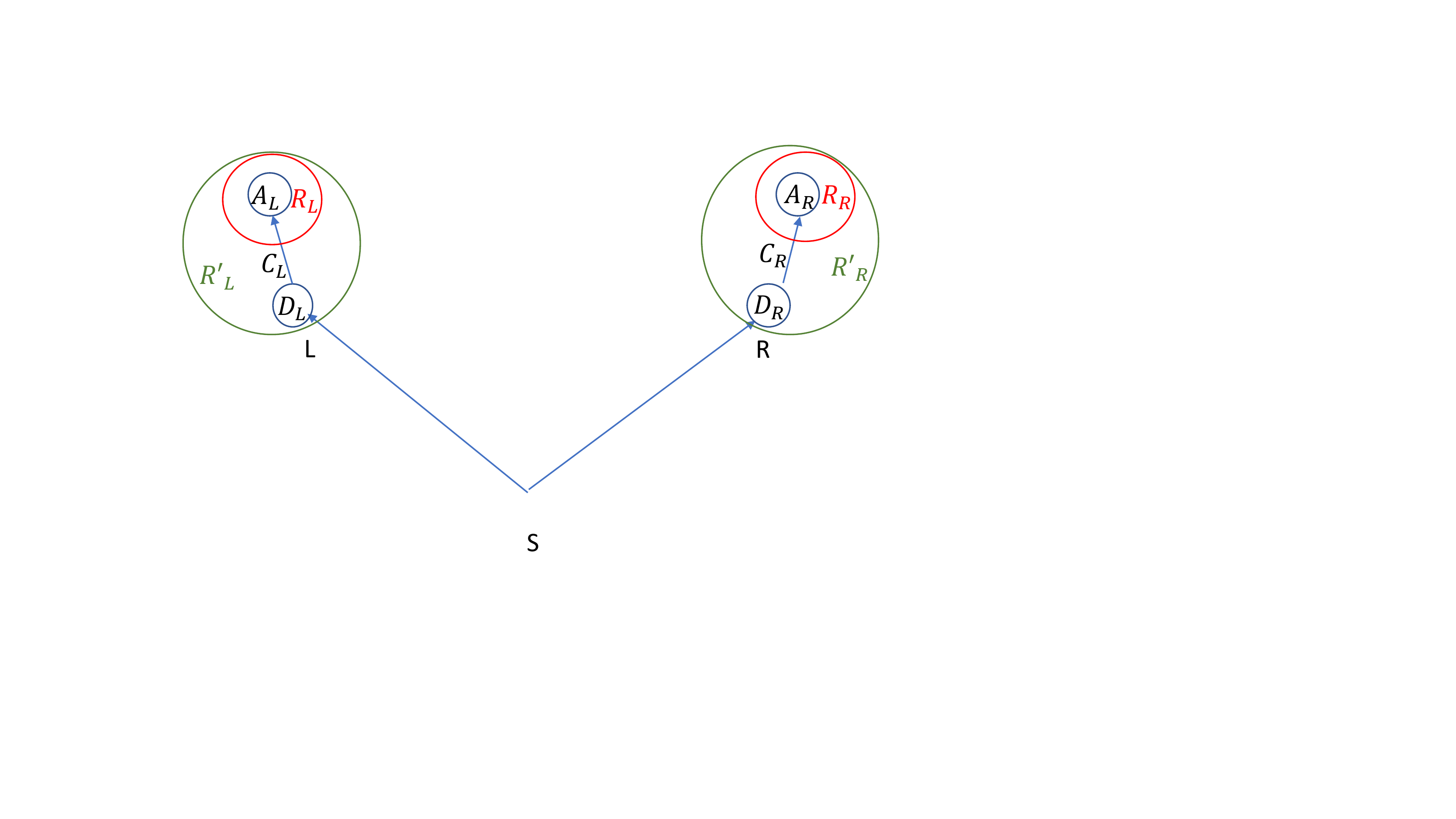} 
\caption{Schematic description of an experiment designed to
close some version of the collapse locality loophole. The
detector readings are communicated via channels to apparatus
which, in an appropriate sense, amplify them to ensure that
collapses are induced.   The regions $R_L$ and $R_R$ are
space-like separated.}
\label{four}
\end{figure}

We can schematically summarize a class of experiments to test 
the collapse locality loophole, including that of Salart et al.,
by Fig. \ref{four}. 
A standard Bell experiment takes place, with a source
$S$ and detectors $D_L$ and $D_R$ whose measurement settings
are adjustable: they are adjusted by $A$ and $B$ (or by
their appropriately programmed devices) to choose 
random or pseudo-random settings for each run.   
The measurement outcomes for each run are propagated
via channels $C_L$ and $C_R$.   They are then amplified 
by apparatus $A_L$ and $A_R$, with the amplification processes
taking place in space-like separated regions $R_L$ and $R_R$.
The full process on each wing $X$, from entering the
detector $D_X$ to the completion of amplification by $A_x$,
takes place in a region which we denote by $R'_X$. 
In the Salart et al. experiments, the regions $R'_L$ and
$R'_R$ were also space-like separated.

Any given such experiment is designed to test the loophole
given the hypothesis that 
collapses take place because of amplification by the apparatus $A_L$ and
$A_R$ and within the regions $R_L$ and $R_R$. 
For an interesting experiment, this hypothesis must be based on well motivated theoretical
ideas, which must also imply that collapse does
{\em not} typically take place either in the detectors $D_X$ (where
$X = L$ or $R$) or the channels
$C_X$.  If it did, then the apparatus would be unnecessary.
Moreover, in this case the experiment might or might not be testing the loophole, depending
whether the regions during which the collapses typically take place are
spacelike separated.   
In particular, if the regions $R'_L$ and $R'_R$ are not space-like
separated, then
typical pairs of collapses may not necessarily be space-like separated.
On the hypothesis that the collapses take place during 
propagation in the channels, then whether they are space-like separated depends on the collapse
duration and the channel configurations.   (We do not 
necessarily assume the channels propagate light-like signals.)
On the hypothesis that the collapses take place within the detectors, and the 
detector measurements take place in space-like separated
regions (as in standard Bell experiments), then standard
Bell experiments would already have closed the loophole. 

At first sight, it may seem that the channels should necessarily be
very short. 
One might also think that the entire regions $R'_L$ and $R'_R$, 
including all the processes from the systems entering the
detectors to the completion of the amplification, should necessarily be  
space-like separated.   
Both of these conditions were satisfied in the Salart et
al. experiment.  
Indeed, the space-like separation of $R'_L$ and $R'_R$ {\it is}
required in order to close the extended version of the collapse locality loophole,
which effectively exploits the locality loophole, so that local
hidden variables in the causal future of $D_X$ may be significantly influenced
by the measurement settings of these detectors. 
However, neither of the above conditions is 
required to close the essential collapse locality loophole, in which
local hidden variables are significantly influenced only by (and in
the causal futures of) collapse events.    
On the hypothesis that collapses are caused by interaction with $A_X$ 
and take place within the regions $R_X$
the detectors $D_X$ and channels $C_X$ effectively form part of 
the entangled system measured in an extended Bell experiment.   
To test the essential collapse locality loophole, all that matters is that $R_L$ and 
$R_R$ are space-like separated.

\begin{figure}[h]
\centering
\includegraphics[width=\linewidth, height=10cm]{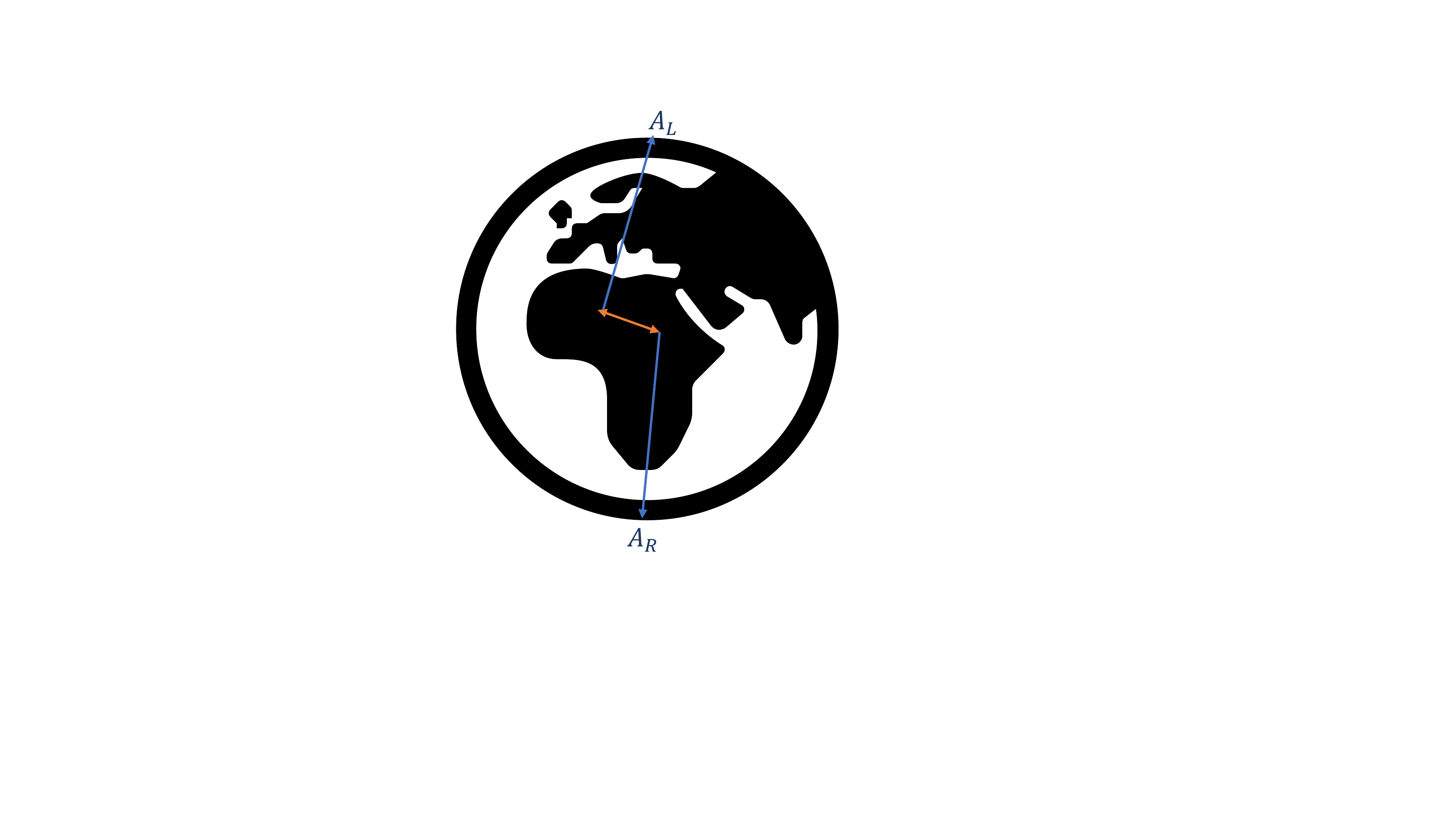} 
\caption{A long range terrestrial experiment designed to
test the essential collapse locality loophole. The
detector readings from wings of a short range Bell experiment are
communicated to amplifying apparatus at antipodal points. 
By introducing delays if necessary, they are input into the apparatus nearly simultaneously
in rest frame, so as to maximize the collapse time for which
space-like separated collapses would ensure.}
\label{five}
\end{figure}

\begin{figure}[h]
\centering
\includegraphics[width=\linewidth, height=10cm]{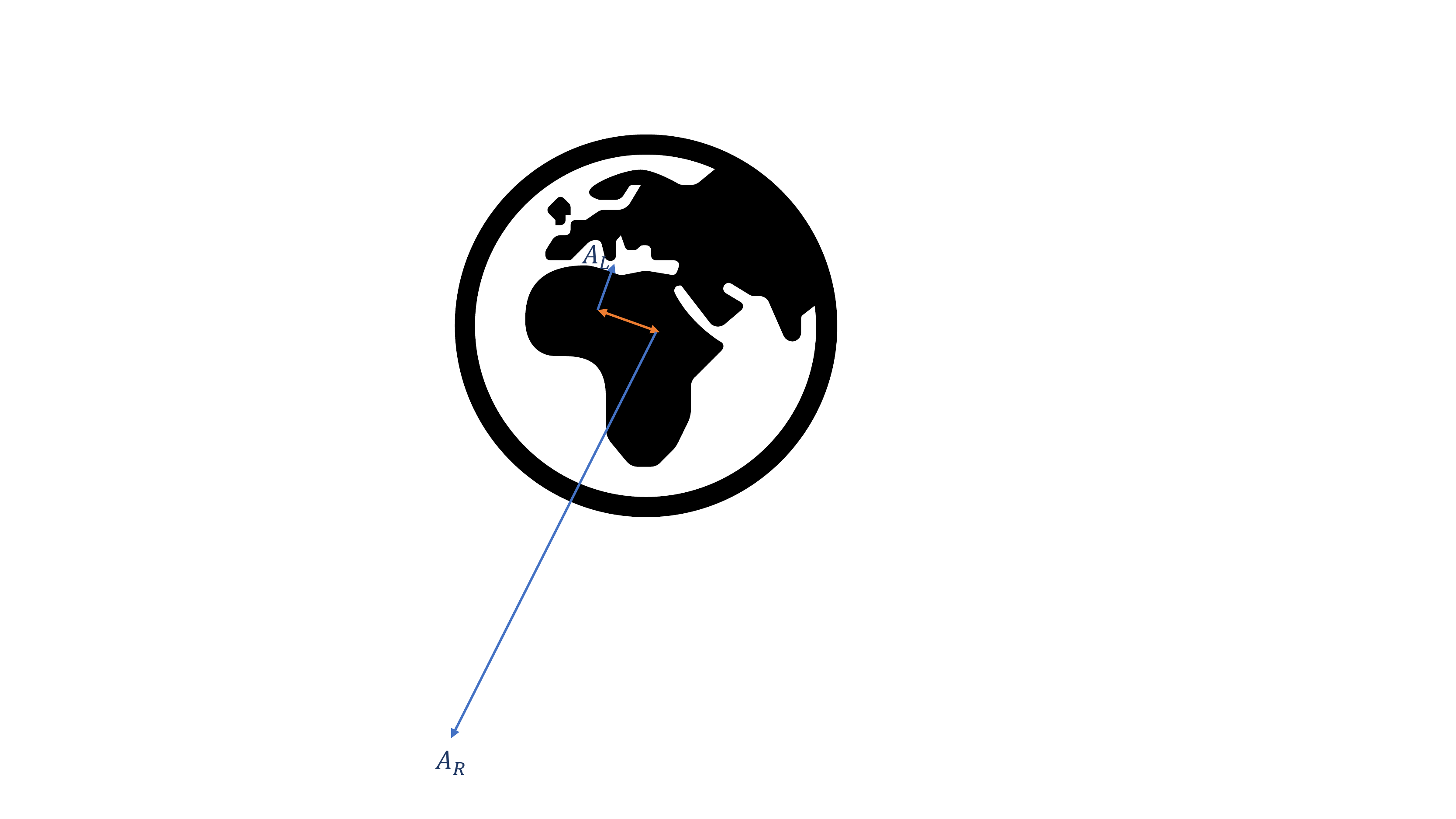} 
\caption{A partly space-based version of the previous experiment.
One signal is sent to an apparatus on a space-based laboratory, while the other 
goes to an apparatus on a ground station.   
To test the Wigner version of the essential collapse locality loophole, the
apparatus may include human observers.}
\label{six}
\end{figure}

This gives us considerable freedom in designing experiments.
In particular, the channels may be slow and long, compared
to the other experimental parameters.  
Moreover, the (by hypothesis uncollapsed) detector ``measurements''  
need not even be space-like separated.  
For example, several interesting versions of the essential collapse
locality loophole could be tested by a Bell experiment with nearly adjacent detectors, with outcomes
propagated to antipodal points on the Earth by fibre optic links or 
radio signals, followed by suitable synchronised amplification at the antipodes
(see Fig. \ref{five}).
Stronger tests still could be carried out by experiments
in which one or both amplifying devices are located in space
(see Fig. \ref{six}).
Again, the Bell experiment detectors need not be widely 
separated, and could both be on Earth, so no long range
controlled distribution of entangled photons is required.

These experiments have the unusual feature that they are
based on the assumption that they generate long range entanglement
of subsystems that include some degrees of freedom (such 
as electrical signals) normally treated
as classical.    This assumption follows from some specified
collapse model or hypothesis.   It is not required that the
entangled 
subsystems be precisely identified or isolated:
decohering interactions with the environment
are not necessarily problematic.  However, the
experiment needs to ensure that any such decohering interactions
are not of the form that, according to the relevant
collapse hypothesis, leads to collapse in the past 
of the apparatus.   For example, humans peeking
at the detector output data before they arrive
at the final apparatus could invalidate the 
tests of the Wigner version of the collapse 
locality loophole described below. 
   
To illustrate some of the power of this technique, note
that the Salart et al. experiment could be modified 
by adding terrestrial channels to antipodal points, 
with synchronization of the propagation of the
signals through antipodal piezocrystals.  
This would allow a separation of the Earth's
diameter, i.e. $\approx 1.24 \times 10^4 {\rm km} \approx 
40 {\rm ms} \times c$,
an improvement by a factor of $\approx 10^3$,
allowing a corresponding margin against the considerable theoretical
uncertainties 
in the collapse time estimates for Diosi and Penrose's proposals and
for other collapse models . 
The increased time length would also allow scope for 
larger and slower piezocrystals and for other 
means of generating distinct mass distributions in
response to signals. 
It is an interesting
challenge for technological ingenuity to identify the largest scale
event that can safely be created, conditioned on a particular signal,
within $\approx 40{\rm ms}$.  As well as piezocrystals, such an event 
could involve motors, triggered springs, and/or controlled explosions, for
example.   Space-based experiments give even longer time intervals
and correspondingly more scope (though for large separations perhaps less critical need) for
ingenuity.

A test of the essential collapse locality loophole based on Wigner's hypothesis 
is also relatively practical with this technique, requiring 
only one astronaut who need carry no specialized equipment.   
Typical human perception or reaction times
of $\approx 100-200 {\rm ms}$ require only that one human participant
is at least $\approx 2-5 \times 10^4 {\rm km}$ from Earth, assuming
that the other is on the opposite side of the Earth; a medium
Earth orbit falls within this range.

Wigner's hypothesis, of course, proposes a direct 
link between consciousness and objective collapse.
It is generally disfavoured, because (a) many physicists 
think that consciousness is weakly emergent from  (i.e.
in principle entirely explicable by) known physics,
(b) many physicists think that quantum physics is 
very plausibly complete, i.e. that there is no
quantum measurement problem and no need to 
add anything to the quantum description of reality,
(c) even among those who believe there is a hard problem
of consciousness and a quantum reality problem, 
most find it hard to see how Wigner's hypothesis
can fit into a plausibly attractive underlying theory. 
Still, many might agree that these are very deep
questions, that our best present theories may yet need
radical revision, our best present intuitions are
not necessarily a very good guide, and experiments
would be worth carrying out for those reasons alone.   

One can also motivate these experiments via dynamical
collapse models.   We have already noted that lower bounds on
the mass-dependent CSL and GRW collapse model parameters
can only be justified by analysing human conscious perceptions.
This also applies to other collapse models that 
could be considered alongside CSL and GRW. 
Ultimately, the only firm reason for believing that a
well motivated collapse model must define a definite
measurement event is that humans would definitely be conscious
of the measurement outcome if they observed it.  
The simplest (and perhaps only) way of ensuring this must be the case for 
all potentially interesting collapse models is to directly involve a human observer in 
the tests.  
This gives quite a strong motivation for tests of the essential collapse locality
loophole directly involving human observers, independent of Wigner's
hypothesis. 

\section{Conclusions}

Collapse hypotheses can be motivated as solutions to the 
quantum reality (or measurement) problem, as alternative routes to
unifying quantum theory and gravity without necessarily quantising
gravity in any standard sense, or even as speculative ways of
connnecting consciousness and physics.  All of these motivations
(which may also be combined in various ways)
are questionable, but all have thoughtful proponents.      
If collapses are objective, it is quite plausible that they
are typically well localized events, and indeed this is 
a feature of some explicit collapse
models. 

Bell non-locality is not necessarily connected with or problematic
for any of these motivations, and it seems pretty likely that it is 
a fundamental fact about nature; this is certainly the 
straightforward explanation of Bell experiments to date.
However the problems of quantum theory, the difficulty in
unifying quantum theory and gravity and the mystery of 
consciousness all counsel a little humility: it is 
still possible that we understand nature much less well than
we imagine.  And a demonstrable failure of Bell nonlocality (despite
appearances) would radically alter the theoretical landscape,
particularly in connecting quantum theory and gravity. 
All of this motivates testing Bell nonlocality as thoroughly
as possible, particularly since Bell experiments are also interesting technological 
and experimental challenges, with spin-offs in applied physics, and
relatively inexpensive.   

Causal quantum theory \cite{kent2005causal,kent2018testing} is
an explicit alternative to quantum theory that exploits the essential
collapse locality loophole.   
However, the loophole may be exploited in other ways.
Tests that could refute causal quantum theory\cite{kent2018testing}
(based on some specific localized collapse hypothesis)
thus would not necessarily close the essential collapse locality
loophole (based on the same hypothesis).   

There is also a cryptographic motivation for considering Bell
experiment loopholes and how to close them.
It is often crucial for future
users of quantum cryptography and quantum communication systems to
guard against eavesdropping or cheating by testing that states involving
allegedly entangled separated subsystems genuinely are entangled
states of the correct form.  In principle, Bell experiments 
are certifications of entanglement.  However, for users working
with untrusted devices, in principle, every unclosed Bell experiment loophole 
gives adversaries a cheating strategy.   In particular, the collapse 
locality loophole focuses attention on whether users know for sure
when and where their Bell measurement outcomes are actually generated. 

For all these reasons, we hope and expect that our techniques will
be exploited and extended. 
We have focussed on experiments aimed at 
closing the essential collapse locality
loophole, assuming that nature does not exploit collapse locality
in combination with other loopholes.   This follows in the tradition of many
significant Bell experiments that addressed
either the detector efficiency or the locality loophole
but not both.  Nonetheless, although we are not aware of any
interesting alternative to quantum theory that exploits the extended
collapse locality loophole, or other combinations of loopholes
including collapse locality, we cannot exclude
the possibility that one might be devised. 
It would thus also be 
desirable to design further, even more definitive, experiments that could ultimately 
simultaneously close the locality, collapse locality,
detector efficiency and other loopholes.

\vskip10pt
\begin{acknowledgments}
I am very grateful to Philip Pearle for many helpful suggestions and 
for estimates of collapse rates in the CSL and GRW models. 

This work was partially 
supported by FQXi and by
Perimeter Institute for Theoretical Physics. Research at Perimeter
Institute is supported by the Government of Canada through Industry
Canada and by the Province of Ontario through the Ministry of
Research and Innovation.
\end{acknowledgments}

\bibliographystyle{unsrtnat}
\bibliography{collapselocexpt}{}
\end{document}